\documentclass[preprint, 12pt]{elsarticle}
\usepackage[top=2.5cm,bottom=2.5cm,left=2.5cm,right=2.5cm]{geometry}
\usepackage[ruled,linesnumbered]{algorithm2e}
\usepackage{mathtools}
\usepackage{multicol}
\usepackage{multirow}
\usepackage{booktabs,tabularx,dcolumn}
\usepackage{xcolor}
\usepackage{amsfonts}
\usepackage{setspace}

\newcommand{\now}{\text{now}}

\newcommand{\stl}{\text{settle}}
\newcommand{\lstl}{\text{last settled}}
\newcommand{\dlv}{\text{deliver}}
\newcommand{\bill}{\textit{bill}}

\newcommand{\start}{\text{start}}
\newcommand{\stopp}{\text{stop}}
\usepackage{xcolor}

\journal{Energy and AI}

\begin{document}

\begin{frontmatter}
%
\title{Decentralized Coordination of Distributed Energy Resources through Local Energy Markets and Deep Reinforcement Learning}
%
%
%

\author[inst1]{Daniel C. May}
            
\author[inst2]{Matthew Taylor}

\author[inst1,inst3]{Petr Musilek\corref{cor1}}
\ead{pmusilek@ualberta.ca}
\cortext[cor1]{Corresponding author}

\affiliation[inst1]{organization={Electrical and Computer Engineering, University of Alberta},
            city={Edmonton},
            postcode={T6G 2R3}, 
            state={AB},
            country={Canada}}

\affiliation[inst2]{organization={Computing Science, University of Alberta \& Alberta Machine Intelligence Institute},
            city={Edmonton},
            postcode={T6G 2R3}, 
            state={AB},
            country={Canada}}

\affiliation[inst3]{organization={Applied Cybernetics, University of Hradec Králové},
            addressline={500 03}, 
            city={Hradec Králové},
            country={Czech Republic}}

\begin{abstract}
As the energy landscape evolves toward sustainability, the accelerating integration of distributed energy resources poses challenges to the operability and reliability of the electricity grid. One significant aspect of this issue is the notable increase in net load variability at the grid edge.

Transactive energy, implemented through local energy markets, has recently garnered attention as a promising solution to address the grid challenges in the form of decentralized, indirect demand response on a community level. Model-free control approaches, such as deep reinforcement learning (DRL), show promise for the decentralized automation of participation within this context. Existing studies at the intersection of transactive energy and model-free control primarily focus on socioeconomic and self-consumption metrics, overlooking the crucial goal of reducing community-level net load variability.

This study addresses this gap by training a set of deep reinforcement learning agents to automate end-user participation in an economy-driven, autonomous local energy market (ALEX). In this setting, agents do not share information and only prioritize individual bill optimization. The study unveils a clear correlation between bill reduction and reduced net load variability. The impact on net load variability is assessed over various time horizons using metrics such as ramping rate, daily and monthly load factor, as well as daily average and total peak export and import on an open-source dataset.

To examine the performance of the proposed DRL method, its agents are benchmarked against a near-optimal dynamic programming method, using a no-control scenario as the baseline. The dynamic programming benchmark reduces average daily import, export, and peak demand by 22.05\%, 83.92\%, and 24.09\%, respectively. The RL agents demonstrate comparable or superior performance, with improvements of 21.93\%, 84.46\%, and 27.02\% on these metrics. This demonstrates that DRL can be effectively employed for such tasks, as they are inherently scalable with near-optimal performance in decentralized grid management.
\end{abstract}

\begin{keyword}
Reinforcement Learning, Deep Reinforcement Learning, Distributed Energy Resources, Local Energy Markets, Demand Response, Distributed Energy Resource Management, Transactive Energy
\end{keyword}

\end{frontmatter}

%



\section{Introduction}\label{sec: Intoduction}

Progress towards sustainable energy utilization is crucial for addressing climate change. In this context, the convergence of technological advances and lagging regulatory frameworks has precipitated the rapid adoption of distributed energy resources (DERs), reshaping the dynamics of the grid edge where electricity end-users reside~\cite{IEA}. Consequently, the variability of the net load at the grid edge is rapidly increasing. The term variability encompasses the composite effects of intermittency and other net load volatilities, such as those caused by electric vehicle charging. This marked increase amplifies the challenges associated with ensuring the reliability and efficiency of grid operations~\cite{KOK_Landscape_Report, OConnell_Landscape_Report}. This drives the transition to the Smart Grid, which operates in a decentralized and autonomous manner to maintain and possibly enhance the operability of the electricity grid.

To address these challenges, the research community has been actively exploring demand response (DR) methodologies. Broadly speaking, DR techniques leverage various signals to modulate end-user net load demand, supporting electrical grid efficiency and reliability. These signals encompass both direct control commands to assets and incentive mechanisms intended to influence end-user behavior, thus delineating between direct and indirect DR. Notably, the key hurdles in indirect DR lie in aligning the interests of grid stakeholders and electricity end-users through appropriate incentive structures and subsequently ensuring sufficient participation to achieve the desired effect~\cite{CHEN2017, FedEnergyRegulatoryCommission_DRReport, Tushar_LEM_Review}. 

Traditionally, price schedule-based approaches have been predominant in indirect DR. Model predictive control (MPC) frameworks have been leveraged to determine such price schedules on a dynamic, state-dependent basis. These approaches rely on behavioral models to form a forecast and then attempt to optimize load demand over a future time horizon. However, their inherent reliance on expert knowledge, high time complexity, and bias toward centralized information processing may impede their efficacy in addressing the rapid and disparate changes observed at the grid edge.

In response to the challenges faced by these scheduling-based methods, transactive energy (TE) has emerged as a compelling alternative. TE, defined as ``the use of a combination of economic and control techniques to improve grid reliability and efficiency" by the GridWise Architecture Council~\cite{GridWise}, aligns well with the Smart Grid ethos, emphasizing the market as a decentralized delivery mechanism for incentive signals~\cite{CHEN2017}.

Recent literature has highlighted the concept of Local Energy Markets (LEMs) as a viable path to implement TE within geographically constrained communities at the grid edge. Mengelkamp et al. define LEM as ``a geographically distinct and socially close community of residential prosumers and consumers who can trade locally produced electricity within their community. For this, all actors must have access to a local market platform on which (buy) bids and (ask) offers for local electricity are matched"~\cite{Mengelkamp2019Review}. LEMs allow for the delivery of real-time incentive signals to electricity end-users, providing the necessary granularity and immediacy within a decentralizable framework.

The surveys of completed DR pilot studies confirm that automation is necessary to facilitate sufficient levels of participation~\cite{CHEN2017, FedEnergyRegulatoryCommission_DRReport, Tushar_LEM_Review}. While MPC is entrenched in the general DR literature for automation, model-free approaches such as deep reinforcement learning (DRL) present a promising paradigm better suited to tackle the challenges faced at the grid edge. Initially inspired by high-level performance showcases of DRL in games~\cite{DQN, Alpha0, OpenAi5}, this notion is reinforced by the success of DRL in fields like robotics~\cite{DRL_in_ROBOTICS} and process control~\cite{DRL_in_ProcessControl}. Moreover, it is supported by a growing body of research applying DRL to the electricity grid~\cite{VCReview, perera2021_DRL_InEnergy, cao2020_DRLInPower_Review}. 

The main advantages of DRL are model independence, inherent adaptability, scalability, and the ability to learn online~\cite{DRL_in_ProcessControl}. In general, these benefits must be weighted against the potential shortcomings of training duration, data requirements, and sample efficiency. However, these issues are not significant in the transactive market environment, which in itself provides a large amount of data stemming from continuous interactions of market participants.

Within this context, recent studies have explored automating end-user participation and DER management in LEMs~\cite{Ye_LEM_1, Ye_LEM_2, Xu_LEM_Article, Bose_Mengelkamp_LEM_Article, Chen_LEMDRL_Article, Zhou_LEM_Article, Zang_LEM_Article}, predominantly through agents trained to optimize end-user bills via load-shifting capacities. Some studies demonstrate the reduction of net community energy consumption~\cite{Xu_LEM_Article, Chen_LEMDRL_Article}, while others investigate the provision of flexibility services~\cite{Ye_LEM_2}. However, to the best of the authors' knowledge, there are no other studies demonstrating the reduction of community-level load variability through the automation of LEMs using DRL.

Such a conclusive demonstration is not trivial. Despite the intention of LEMs to align the interests of end-users with the objectives of grid stakeholders, it is crucial to recognize that incentivized behavior may not automatically translate into reduced variability or enhanced power quality at the local level~\cite{Kiedanski_LEM_Lit_Issues, Papadaskalopoulos_LEM_Lit_Issues}. Similarly, the intricate interplay between LEM design and participant automation may yield unforeseen outcomes~\cite{Mengelkamp_NoDRL_LEM_Article}, a phenomenon commonly observed when automating complex systems using DRL~\cite{openAi_HideAndSeek}.

Building on prior studies by Zhang et al.~\cite{ALEXV1} and May et al.~\cite{ALEXV2}, this article addresses this research gap by training independent agents to automate end-user participation in LEMs and the utilization of DERs. The main contributions of this study are as follows:
\begin{itemize}
    \item Demonstrates a reduction in community net load variability, showing that selfish bill optimization by agents within LEMs can lead to emergent community-level benefits.
    \item Conducts performance evaluation on an open-source dataset, enhancing benchmarking and future comparability of the study's results.
    \item Benchmarks DRL agents against several baselines, showing that the trained DRL agents achieve performance equivalent to a near-optimal dynamic programming benchmark.
\end{itemize}


Subsequent sections of this article delve into related work and background in Section~\ref{sec: Related Lit and Background}, methodology for training DRL agents, evaluation and benchmarking procedures in Section~\ref{sec: Methodology}, a comprehensive discussion of simulation results in Section~\ref{sec: Discussion}, and conclude with a brief summary and avenues for future research in Section~\ref{sec: Conclusion}.

\section{Related Work and Background}\label{sec: Related Lit and Background}

Subsection~\ref{subsec: Related Lit} briefly reviews related literature and establishes a notable research gap: the lack of a well-benchmarked demonstration of variability reductions within an economy-driven LEM, emerging from selfish end-user bill minimization that DRL agents automate.
Subsection~\ref{subsec: ALEX} introduces the LEM design that forms the foundation of this study.
Subsection~\ref{subsec: DRL} overviews reinforcement learning and proximal policy optimization, the base DRL algorithm employed within this article.

\subsection{Related Literature}\label{subsec: Related Lit}

The application of DRL in DR, and for the electricity grid in general, has garnered significant attention in recent years~\cite{VCReview, cao2020_DRLInPower_Review, perera2021_DRL_InEnergy}. Studies exploring the distributed coordination of DERs through DR mechanisms outside of LEM, such as those by Chung et al.~\cite{Chung_NoLEM_Article}, Zhang et al.~\cite{Zhang_NoLEM_Article}, and Nweye et al.~\cite{Nweye_DR_DRL_Article_citylearn}, tend to optimize for composite rewards and incorporate community-level metrics related to grid stability or variability, following a direct optimization approach.  

Concurrently, there has been a surge in literature investigating LEMs. Mengelkamp et al.~\cite{Mengelkamp2019Review}, Capper et al.~\cite{CAPPER_LEM_Review}, and Tushar et al.~\cite{Tushar_LEM_Review} provide comprehensive insights into the evolving LEM ecosystem. In general, this field tends to focus on the socioeconomic performance of the proposed system, while DR aspects are only narrowly discussed, and performance benchmarking tends to be restricted.

For instance, Liu et al.~\cite{Liu_LEM_Article} propose a LEM-like mechanism, using pricing based on the supply-demand ratio to coordinate energy flow between microgrids, leveraging MPC for automation.
Similarly, Lezama et al.~\cite{Lezama_LEM} explore LEMs from a grid integration perspective, focusing on socioeconomic performance.
Ghorani et al.~\cite{Ghorani_NoDRL_LEM_Article} develop bidding models for risk-neutral and risk-averse LEM agents, evaluating their socioeconomic efficacy under various market designs.
Meanwhile, Mengelkamp et al.~\cite{Mengelkamp_NoDRL_LEM_Article} investigate different market designs using heuristic agents, focusing on socioeconomic metrics.
A burgeoning body of research emphasizes the automation of LEM participation
through DRL.
Xu et al.~\cite{Xu_LEM_Article} employ a MARL Q-learning algorithm to automate participation in a LEM that communicates a pricing schedule based on a supply and demand forecast.
Zhou et al.~\cite{Zhou_LEM_Article} propose an economy-driven LEM pricing mechanism, optimizing participant bidding via a combination of Q-learning and fuzzy logic.
Similarly, Zang et al.~\cite{Zang_LEM_Article} train end-user agents to interact with community-level batteries within LEMs.

As Mengelkamp et al.~\cite{Mengelkamp_NoDRL_LEM_Article} highlight, the integration of LEMs and automated participation presents complex challenges and potentially unforeseen consequences due to the emergent, intricate system dynamics.
Investigations by Kiedanski et al.~\cite{Kiedanski_LEM_Lit_Issues} and Papadaskalopoulose et al.~\cite{Papadaskalopoulos_LEM_Lit_Issues} demonstrate that increases in socioeconomic performance in such settings may not directly translate to improved grid performance in terms of reducing variability or improving power quality.

To address this issue, some studies incorporate electricity grid performance metrics into the LEM's pricing mechanism or the agent's reward function, diverging from the original purely economic focus of LEMs and adopting a direct optimization approach. For example, Chen et al.\cite{Chen_LEMDRL_Article} investigate microgrid trading in the context of LEMs, employing a reward function with explicit constraints. Their findings demonstrate that this approach increases self-sufficiency compared to expert-designed heuristics and random action agents in benchmarking experiments. Similarly, Ye et al.~\cite{Ye_LEM_2} explore the use of LEMs to provide flexibility services. Their contribution stands out by benchmarking against a near-optimal MPC baseline, establishing a reasonable upper performance limit. However, even such contributions do not evaluate their agents' performance on variability-related metrics for which the agents do not explicitly optimize.

The principal promise of LEM, and, in a more general sense, TE, lies in the notion that a well-designed market mechanism should incentivize a broad range of beneficial behaviors. The underlying ambition is to achieve this without explicitly tying the market's cost function to these outcomes, enabling agile and robust decentralization by avoiding the need for expensive real-time computation of an expressive set of related metrics. In a sense, optimizing end-user bills should indirectly and emergently reduce net load variability in this setting. Despite the current landscape of contributions, the demonstration of such behavior via an LEM that relies on DRL for automation purposes is still outstanding. This study aims to contribute to closing this research gap.

\subsection{Autonomous Local Energy eXchange}~\label{subsec: ALEX}

The Autonomous Local Energy eXchange (ALEX), initially proposed by Zhang et al.~\cite{ALEXV1}, serves as an LEM for a community denoted as $B$, where individual buildings $b \in B$ participate in energy trading facilitated by a round-based, futures-blind double auction settlement mechanism. In the context of a round-based futures market, trading occurs in predefined time intervals. A futures market accepts bids and asks for a future settlement timestep $t_{\stl}$, to be submitted at the current time step $t_{\now}$. Settlements are then executed at a subsequent time step $t_{\dlv}$, with $t_{\dlv} > t_{\stl} > t_{\now}$. In such a blind double auction market, each building interacts with the market without awareness of other buildings' activities.
 
In ALEX, market participants do not share information and instead selfishly optimize their individual electricity bills. The building's electricity bill consists of two main components: the market bill and the grid bill. The market bill includes the cumulative settlement cost, determined by pairing each settlement price with its corresponding quantity for the building. In contrast, the grid bill covers the residual amount required to meet the household's energy demand, billed at the prevailing grid rate selling or buying price, depending on the current net-billing scenario. A profitability margin between the grid rate selling and buying prices serves as an incentive for LEM utilization. This means that any exchange over the LEM presents a favorable scenario. Achieving this could involve leveraging tracked greenhouse gas emission savings or partial fee offsets~\cite{CAPPER_LEM_Review, Dudjak_LEM_Review}.

Zhang et al.~\cite{ALEXV1} delve into the essential properties required for ALEX's settlement mechanism to incentivize RL agents to learn pricing in correlation with the settlement timestep $t_{\stl}$ supply and demand ratios. Formulating ALEX as a competitive mixed-form stochastic game suggests the existence of at least one Nash equilibrium.\footnote{Note that this game is not zero sum --- one agent does not succeed only at the expense of another.} This insight facilitates the identification of a market mechanism possessing the desired properties through experiments that employ tabular Q-learning bandits under varied but fixed supply and demand ratios. Subsequent experiments deduce a market price function based on the current supply and demand ratio. A follow-up study by Zhang and Musilek~\cite{Zhang_ALEX_MDPI} investigates a system incorporating a communal battery energy storage system (BESS) controlled by an expert-designed heuristic. The study demonstrates efficacy in avoiding violations of voltage-frequency constraints on a test circuit.

May and Musilek~\cite{ALEXV2} further examines ALEX as a DR system. The authors simulate a group of near-optimal, rational actors on ALEX using an iterative best-response and dynamic programming algorithm. Their performance is then compared against several baselines. The identified policies reveal emergent community-level coordination of DERs, driven by incentives within the LEM. Remarkably, this coordination occurs even though each participant accesses only building-level information and selfishly optimizes their electricity bills. Consequently, these policies consistently outperform the benchmark building-level DR system across various community net-load-related metrics measuring net load variability at the community level. While this agent behavior shows promise, it is important to note that it is generated using a pure search approach that relies on a perfect forecast of end-user generation and demand.

This study aims to extend these results by training a set of DRL agents on the equivalent task without access to perfect forecasts, yet achievening a comparable level of variability reduction. This would effectively address the research gap identified in subsection~\ref{subsec: Related Lit}.

\subsection{Reinforcement Learning and Markov Decision Processes}\label{subsec: DRL}

\begin{figure}[ht]
    \centering
    \includegraphics[width=0.5\textwidth]{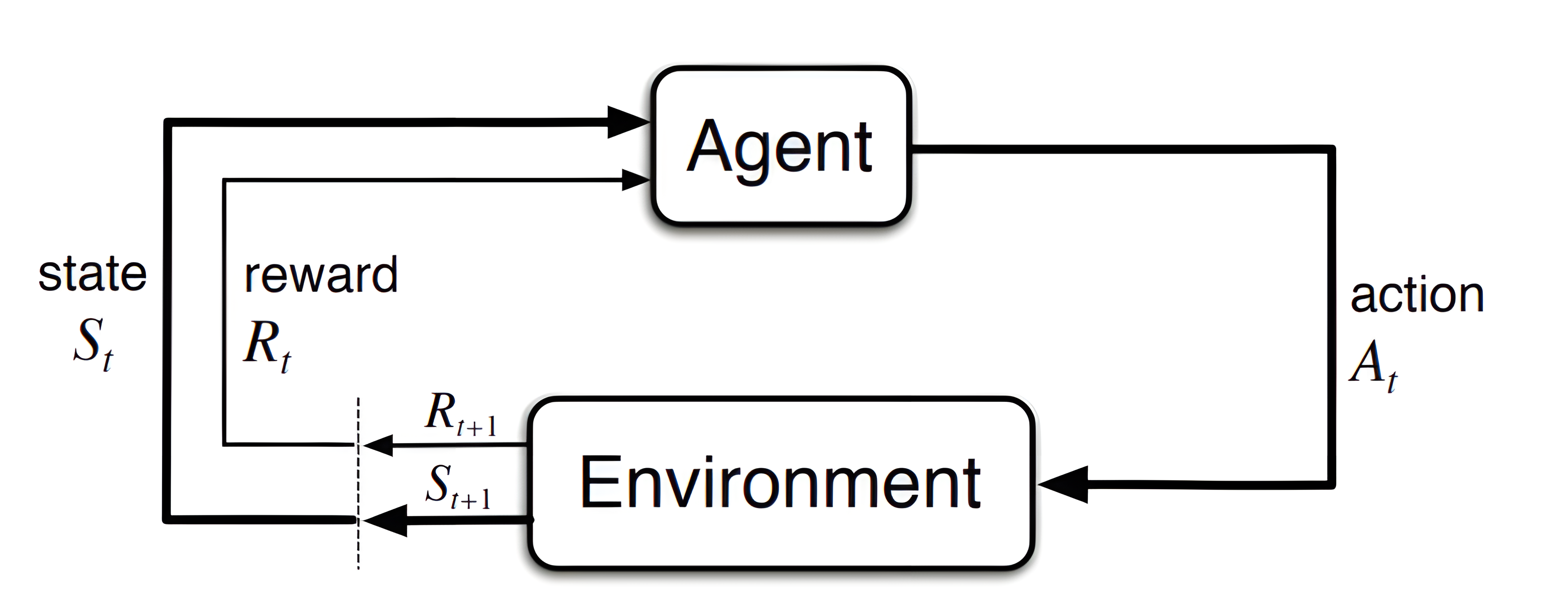}
    \caption{Agent to environment interaction diagram, taken from Sutton \& Barto~\cite{Sutton}.}
    \label{fig: AgentEnvInteraction}
\end{figure}

Reinforcement Learning (RL) is a machine learning framework for optimizing agent interactions with an environment based on a reward signal, as illustrated in Figure~\ref{fig: AgentEnvInteraction}.

This is formalized through the Markov Decision Process (MDP), represented by the tuple $(S, A, P_a, R_a)$. Here, $S$ denotes the state space, $A$ the action space, $P_a$ the transition probabilities from state $s$ to the next state $s'$ upon taking an action $a$ and $R_a$ is the immediately received reward.
A policy $\pi$ characterizes an agent's behavior through a probabilistic mapping from state $s$ to action $a$, for example in the form of a Gaussian distribution, where the mean $\mu$ and standard deviation $\sigma$ are functions of the state $s$.

RL agents typically operate within a time-discrete MDP, where the time step $t$ represents points along the interaction trajectory, starting at $t=0$ and concluding at $t=T$. The return $G$ signifies the cumulative, discounted future reward,
\begin{equation}
    \label{eq:Return_Alex3_Alex3}
        G_t = \sum_{t=0}^{T} \gamma^t R_{t+1},
\end{equation}
which facilitates the definition of state value 
\begin{equation}
    \label{eq:StateValue_Alex3}
    V_{\pi}(s_t) = \mathbf{E}G_t  \forall \pi,
\end{equation}
and state-action value 
\begin{equation}
    \label{eq:ActionValue}
    Q_{\pi}(s_t, a_t) = \mathbf{E}G_t \forall \pi,
\end{equation}
where $\gamma$ is the discount factor.
The goal is to identify an optimal policy $\pi^*$ which maximizes the expected return $\mathbf{E}G$.

One of the key strengths of RL is its ability to learn optimal policies by directly interacting with the environment. This allows such agents to handle unknown transition dynamics and latent variables without requiring an explicit environment model.
In contrast to other MDP search methods, RL agents iteratively learn the optimal policy $\pi^*$ through temporal difference learning and bootstrapping, adjusting their parameters $\theta$ in response to the reward signal. These parameters are updated via the RL algorithm's loss function, typically using stochastic gradient descent.
RL algorithms fall into two categories: value-based and policy gradient methods. Value-based methods estimate state values $V$ or state-action values $Q$, while policy gradient methods directly learn policies $\pi$ through a policy loss
\begin{equation}
    \label{eq:NaivePGLoss}
    L(\theta) = \mathbb{E}  \left[ \log \pi_{\theta}(a_t, s_t) V_t\right].
\end{equation}

Actor-critic methods build upon policy gradients, reducing variance by relying on a critic to estimate state values $V$ and compute advantages $A$, 
\begin{equation}
    \label{eq:ACPGLoss}
    L(\theta) = \mathbb{E}  \left[ \log \pi_{\theta}(a_t, s_t) A_t\right].
    A_t = V_t - V_{\theta}(s_t).
\end{equation}

Deep Reinforcement Learning (DRL), combining RL with deep neural networks, has become popular for solving complex MDPs~\cite{Alpha0, OpenAi5, DQN}. DRL methods use replay buffers to store agent-environment interactions for multiple mini-batch gradient updates, requiring differentiation between the parameters used for sampling $\theta_{old}$ and the updated parameters $\theta$.

Proximal Policy Optimization (PPO), a popular actor-critic algorithm introduced by Schulman et al.~\cite{Schulman_PPO}, uses a clipped surrogate objective based on the probability ratio $r(\theta)$, comparing the new policy $\pi_{\theta}$ with the old policy $\pi_{\theta_{old}}$. This prevents policy drift and ensures reliable data collection. The actor loss clips the policy ratio $r(\theta)$ within a tolerance parameter $\epsilon$, 
\begin{equation}
    \label{eq:PPOLoss}
    L(\theta) = \mathbb{E} \left[ \min \left(  r(\theta) A_t, clip(r(\theta), 1-\epsilon, 1+\epsilon) A_t\right) \right].
\end{equation}
Most PPO implementations also incorporate generalized advantage estimation (GAE) to reduce the variance of the advantage $A$, as proposed by Schulman et al.~\cite{Schulman_GAE}.

\section{Methodology and Evaluation}\label{sec: Methodology}

This study aims to extend previous contributions~\cite{ALEXV1, ALEXV2} by training DRL agents to autonomously participate in ALEX. The expectation is that these agents will demonstrate a level of emergent community-level variability reduction that is comparable to the near-optimal search method described by May and Musilek~\cite{ALEXV2}, but without relying on a perfect forecast. Such a showcase of variability reduction within a DRL-driven LEM context would address the significant research gap outlined in Subsection~\ref{subsec: Related Lit}.

To achieve this goal, this section formulates ALEX environment as a Markov Decision Process (MDP) in Subsection~\ref{subsec: ALEXMDP}, outlines the DRL algorithm employed for training the agents in Subsection~\ref{subsec: Algorithm}, and elucidates the experimental design in Subsection~\ref{subsec: Experimental Design}. The latter also includes details on evaluation performance metrics and baselines.

\subsection{Autonomous Local Energy eXchange as Markov Decision Process}\label{subsec: ALEXMDP}

The formulation of ALEX as an MDP involves defining the agent's observations $O$, actions $a$, rewards $r$, and policy $\pi$. In comparison to the initial formulation~\cite{ALEXV2}, the approach outlined here incorporates specific adaptations tailored to the nature of ALEX as a futures market. This is crucial, given the constraint that the DRL agents should not rely on future information. Additionally, the formulation accommodates continuous observation and action spaces for the DRL agents.

The individual agent's MDP encapsulates the viewpoint of a single agent within the ALEX environment. Given that participants in ALEX neither share information nor engage in communication, this individual agent MDP is partially observable. This contrasts with the fully deterministic nature of the joint MDP. In this study, the DRL agents must function as fully independent actors, navigating a continuous action and a partially observable, continuous state space. Accordingly, this section adopts this perspective and refers to the state space $S$ as the observation space $O$.

The observation space $O^b$ for an individual agent at timestep $t$ encompasses various continuous variables, including the current net load $E^b_t$, battery state of charge $SoC^b_t$, the average last settlement price $p^{bid}_{t{\lstl}}$, and total bid and ask quantities from the last settlement round $q^{bid}_{t{\lstl}}$ and $q^{ask}_{t_{\lstl}}$, respectively. To capture temporal patterns such as daily and yearly seasonalities, sine and cosine transformations of the current timestep $t$ are incorporated instead of using the raw timestamp.

\begin{equation}
    \label{eq:StateSpace}
    \begin{split}
        O^b_{t} := \left( \right. sin(t)_{year}, cos(t)_{year}, sin(t)_{day}, cos(t)_{day}, \\
         E^{b}_{t}, SoC^{b}_{t}, p_{t_{\lstl}}, q^{bid}_{t_{\lstl}}, q^{ask}_{t_{\lstl}} \left. \right).
    \end{split}
\end{equation}

However, future information, such as net load at settlement time $E^{b}_{t_{\stl}}$, is not included in this observation space. When designing this observation space, preliminary tests suggest the described feature set as a minimum necessary feature set that enables optimal decision-making.

In contrast to the action space proposed by Zhang et al.~\cite{ALEXV1}, the action space $A^b$ for an agent at timestep $t$ exclusively includes the continuous battery action, scheduled for the future settlement time step $t_{\stl}$. This action is constrained by the battery's charge and discharge rates. The determination of bid and ask quantities at settlement time $t_{\stl}$ relies on the residual net load, while bid and ask market conditions dictate prices following the round's closure, guided by the price curve defined by Zhang et al.~\cite{ALEXV1}.

The building policy $\pi_{b}$ is defined as a superposition of two policies: 
\begin{itemize}
    \item The self-sufficiency maximizing, greedy battery policy $\pi_{0}$ consisting only out of $\mu_{0}$.
    \item The agent's learned policy $\pi_{\theta}$ with mean $\mu_{\theta}$ and $\sigma_{\theta}$.
\end{itemize}
resulting in a Gaussian distribution with 
\begin{equation}
    \label{eq:AgentCompositeAction}
    \mu_{b} = \mu_{0} + \mu_{\theta},
\end{equation}
\begin{equation}
    \label{eq:AgentCompositeAction2}
    \sigma_{b} = \sigma_{\theta}.
\end{equation}
This action and agent policy definition offers several distinct advantages, significantly expediting the learning process of the studied DRL agents. The policy $\pi_{0}$ can be computed at settlement time and serves as a reasonable initial heuristic, even though it may be far from the optimal policy. This approach enables more efficient state exploration while mitigating some of the internal environment modeling that the agent has to perform. The constraints to the agent's actions are enforced by squashing $\pi_{b}$ into the appropriate range using $tanh$, similar to how SAC enforces a bounded policy~\cite{SAC}.

The agents are tasked with minimizing their market-bound electricity bill $\bill^{b}$.
The agent's reward function is formulated as the inverse difference between the electricity bill $\bill^{b}_{t_{\stl}}$ and the bill incurred by the self-sufficiency maximizing policy $\pi_{0}$, denoted as $\bill^{b, \pi_{0}}_{t{\stl}}$,  
\begin{equation}
    \label{eq:CompositeBillAdvantage}
    r^{b}_{t}  :=  \bill^{b, \pi_{0}}_{t_{\stl}} - \bill^{b}_{t_{\stl}}.
\end{equation}
Compared to using the naive participant electricity bill $\bill^{b}_{t}$ as a reward signal, this offers a clearer indication of whether the RL agents are learning a helpful policy since a positive reward value means the agents outperform the greedy policy baseline.

\subsection{Shared Experience Recurrent Proximal Policy Optimization}\label{subsec: Algorithm}

In initial exploration, off-policy methods such as Soft Actor-Critic (SAC)~\cite{SAC} performed worse than on-policy methods such as PPO.
The DRL algorihtm of choice for this study is therefore PPO, introduced in Section~\ref{subsec: DRL}. It is a well-established algorithm with numerous existing and well-benchmarked implementations, which provide a reliable foundation for expansion and adaptation.
Compared to value-based methods, PPO is better suited for the continuous action spaces present in the to-be-solved MDP. Furthermore, acting probabilistically can be beneficial in solving mixed-form stochastic games like ALEX.
One of the key reasons for selecting RL over other applicable methods is its ability to handle uncertainty in the environment's dynamics through model-free approaches.

The agents in this study undergo training as independent agents with shared experience~\cite{SharedExperienceAC}. Although each agent acts autonomously and solely accesses building-level information, they aggregate trajectories into a shared replay buffer. During trajectory collection, the actors function as independent copies of the same actor and critic neural network, which is updated from the shared replay buffer. This maintains full independence between agents during rollout but promotes faster convergence. Christianos et al.~\cite{SharedExperienceAC} demonstrated the efficacy of this approach in enhancing performance within complex multi-agent environments when compared to a fully independent learning setup. Observations undergo standardization and mean-shifting, while rewards are solely standardized, following best practices proposed by Schulman et al.~\cite{schulman2016nuts}.

The remaining portion of this section details modifications to the underlying PPO algorithm. A recurrent PPO~\cite{SB3}, using a Long Short-Term Memory (LSTM)~\cite{lstm} hidden layer for both the actor and the critic, is enhanced with recurrent burn-in and initialization, proposed by Kapturowski et al.\cite{Kapturowski_R2D2}. Drawing motivation from the findings of Andrychowicz et al.~\cite{HiddenStateRecalc}, after processing a replay buffer, the new weights $\theta$ are used to recalculate the hidden states of the agent LSTM based on the entire trajectory experienced during the current episode. Both enhancements address the risk of stale or drifted state representations, enhancing the agent's capacity to develop meaningful state representations and a long-term context. Informed by Ilyas et al.~\cite{Ilyas_LookAtDRL_PG} and with the goal of convergence towards a Nash equilibrium, the learning rate is annealed throughout the training. Instead of setting the value for the terminal transition at $T$ to 0, this study takes it from the critic's value prediction, with preliminary tests indicating an accelerated convergence of the critic to a higher explained variance. Furthermore, instead of naively imposing action space boundaries by clipping the Gaussian distribution, the algorithm used in this study employs a squashed Gaussian distribution followed by renormalization, as popularized by soft actor-critic algorithms~\cite{SAC}.

The initialization of the actor's final layer is designed to ensure that the mean $\mu$ exhibits an expected value of 0. This is achieved by sampling the weights and biases of this layer from a uniform distribution between 0.001 and -0.001. In a similar vein, the policy's standard deviation $\sigma$ is initialized very narrowly. This setup enables the agent to commence training based on trajectories collected near the self-sufficiency maximizing policy $\pi_{0}$. This strategy is grounded in the assumption that the optimal policy $\pi^{*}_{\theta}$ is much closer to the self-sufficiency policy $\pi^{0}$ than to a pure random policy. Large deviations from $\pi^0$ are considered highly situational, while smaller deviations are more common. From a task decomposition perspective, the RL agents learn how to load shift to maximize self-sufficiency, an internally focused task, and then proceed to learn how to leverage the market, an externally focused task. Hence, this practice aims to bias the agents to first learn how to load shift and then learn how to utilize the market. Both adjustments contribute to notable improvements in convergence for the studied task as identified in preliminary tests.

Additional auxiliary information such as the computational resources used, runtimes, 
hyperparameter values, and hyperparameter tuning,
please refer to \ref{Appendix: Learning Curves}.

\subsection{Experimental Design} \label{subsec: Experimental Design}
The DRL agents are trained and evaluated on the CityLearn2022 dataset~\cite{CityLearn2022Data}. The open-source nature of this dataset enables subsequent studies to directly benchmark against this contribution across a diverse range of DR applications. This dataset provides a year of hourly data for 17 smart community buildings, featuring time series of energy demand, photovoltaic generation, and BESS performance characteristics. 
As is common in studies like this, it is assumed that the larger grid operates as an infinite source and sink, able to absorb and supply all residual load without limitations at both the prosumer and community levels.
For each building, one independently acting agent is trained as outlined in Subsection~\ref{subsec: Algorithm}. Therefore, one episode is defined as a full trajectory over the dataset and lasts 8760 steps, while one run fully trains such a set of 17 agents. For evaluation, the parameter set $\theta$ with the best episodic communal return $G^{B}$ is selected from a run, assuming that this snapshot represents the best-performing equilibrium between agents. This snapshot is updated throughout training when a new best communal return is achieved. From a set of 5 runs, the median performing run is selected for benchmarking purposes.

To assess agent performance, we employ a set of metrics from May and Musilek~\cite{ALEXV2}. All performance metrics in this study are functions of the community net load $E^{B}$, defined as the summation of all building net loads $E^{b}$. The following expressions utilize $n_d$ to denote the number of days in the dataset, $d$ to represent the number of time steps in a day, and $t$ as the current time step. The notations $\max_{\start}^{\stopp}$ and $\min_{\start}^{\stopp}$ denote the maximum and minimum operands over the interval from $\start$ to $\stopp$, respectively. Given the hourly resolution of the dataset used in this study, the conversion from kilowatt-hours (kWh) to kilowatts (kW) is excluded from the notation. The performance metrics encompass:
\begin{itemize}
    \item The average daily imported energy
        \begin{equation}
            \label{eq:AvgDailyImportedEnergy_Alex3}
            \overline{E}_{d, +}=  \frac{1}{n_d} \sum_{d=0}^{n_d} \left( \sum_{t \in d} \max(E^B(t),0) \right)
        \end{equation}
    \item The average exported energy
        \begin{equation}
            \label{eq:AvgDailyExportedEnergy_Alex3}
            \overline{E}_{d, -}=  \frac{-1}{n_d} \sum_{d=0}^{n_d} \left( \sum_{t \in d} \min(E^B(t),0) \right)
        \end{equation}
    \item The average daily peak
        \begin{equation}
        \label{eq:AvgDailyPeak_Alex3}
            \overline{P}_{d,+} = \frac{1}{n_d} \sum_{d=0}^{n_d} \left( \max_{t \in d } E^B(t) \right)
        \end{equation}
    \item The average daily valley
        \begin{equation}
        \label{eq:AvgDailyValley_Alex3}
            \overline{P}_{d,-} = \frac{1}{n_d} \sum_{d=0}^{n_d} \left( \min_{t \in d } E^B(t) \right)
        \end{equation}
    \item The absolute maximum peak
        \begin{equation}
            \label{eq:MaxPeak_Alex3}
            P_{+} = \max_{t=0}^{T} E^B(t)
        \end{equation}
    \item The absolute minimum valley 
        \begin{equation}
            \label{eq:MaxValley_Alex3}
            P_{-} = \min_{t=0}^{T} E^B(t)
        \end{equation}
    \item The average daily ramping rate
        \begin{equation}
        \label{eq:AvgDailyRampRate_Alex3}
            \overline{R}_{d} = \frac{1}{n_d} \sum_{d=0}^{n_d} \left( \sum_{t \in d} |\nabla E^B(t)| \right)
        \end{equation}
    \item The daily load factor complement
        \begin{equation}
        \label{eq:DailyLoadFactor_Alex3}
        1- L_{d} = \frac{1}{n_d} \sum_{d=0}^{n_d} \left( 
                            1 - \frac
                                {\mathrm{mean}_{t \in d} E^B(t)}
                                {\max_{t \in d} E^B(t)}
                        \right)
        \end{equation}
    \item The monthly load factor complement
       \begin{equation}
        \label{eq:MonthlyLoadFactor_Alex3}
        1- L_{m} =  \frac{1}{n_m} \sum_{m=0}^{n_m} \left( 
                            1 - \frac
                                {\mathrm{mean}_{t \in m} E^B(t)}
                                {\max_{t \in m} E^B(t)}
                        \right).
        \end{equation}
\end{itemize}

This comprehensive set of metrics offers insights into the variance of the community net load $E^{B}$ across various time scales. These time scales range from the hourly perspective, as captured by the ramping rate $\overline{R}_{d}$, to daily and monthly perspectives, as captured by the daily and monthly load factors $1- L_{d}$ and $1- L_{m}$. Additionally, the yearly and daily averages of peak load demands and generation values provide valuable information about community energy consumption and infrastructure strains. Importantly, all metrics are formulated so that lower values are preferable. Collectively, these metrics provide a robust framework for assessing the performance of an arbitrary DR system regarding its general impact on net load variability.

To effectively gauge the relative performance of the trained DRL agents, three benchmarks are used:
\begin{itemize}
    \item \textbf{NoDERMS}:
    This baseline corresponds to the default community, where no building exploits its battery storage capacities. It serves as the reference setting and is expected to be easily outperformed by any DR system. 
    \item \textbf{IndividualDERMS}: 
    In this benchmark, each building in the community operates under a net billing strategy. Buildings prioritize self-sufficiency by smoothing building-level peaks and valleys while minimizing the ramping rate~\cite{ALEXV2}. This benchmark serves as a reasonable performance baseline, resembling a well-tuned heuristic system commonly found in current DR applications. Importantly, unlike the proposed DRL agents, this benchmark has access to a perfect forecast.
    \item \textbf{ALEX DP}:
    This benchmark represents a near-optimal policy within a discretized version of ALEX's MDP. It is determined using a dynamic programming search method based on iterative best response and value iteration~\cite{ALEXV2}. Importantly, unlike the proposed DRL agents, this benchmark has access to a perfect forecast.   
\end{itemize}

This experimental design aims to address the research gap identified in Section~\ref{sec: Related Lit and Background} by evaluating ALEX RL's performance based on the outlined metrics. ALEX RL is compared against the NoDERMS baseline, with Individual DERMS serving as a reasonable performance benchmark, and ALEX DP representing a near-optimal solution.

\section{Results and Discussion}\label{sec: Discussion}

This study aims to address a significant research gap highlighted in the background section by demonstrating a reduction in community-level variability of net load facilitated by DRL agents within a LEM.
Towards this objective, this section establishes a clear connection between participant bill reduction and performance metrics within ALEX, assuming converged actors.
Subsequently, a comparative analysis of the DRL agents against benchmarks introduced in the earlier subsection is conducted.

\begin{figure}[h]
    \centering
    \includegraphics[width=0.65\textwidth]{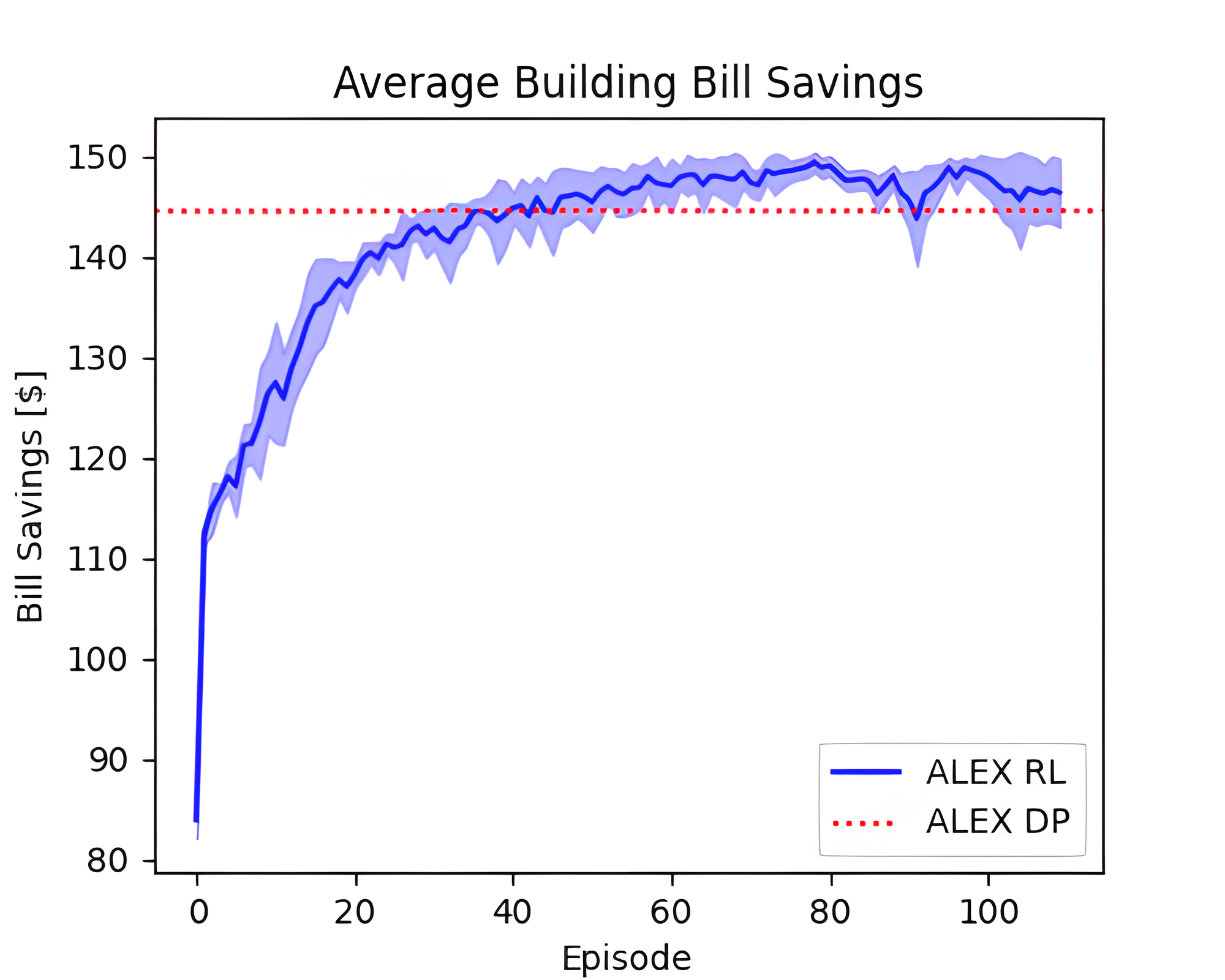}
    \caption{Average participant bill savings comparison between ALEX RL (blue), ALEX DP (red). Shaded areas depict variance bands between a set of 5 ALEX RL runs, trained over 117 episodes.}
    \label{fig: Avg Building Return}
\end{figure}

The training methodology of the agents focuses on their relative improvement compared to the self-sufficiency maximizing policy $\pi^{0}$, as outlined in Section~\ref{subsec: Algorithm}. Convergence behaviors are visually depicted in Figure~\ref{fig: Avg Building Return}, highlighting the average building bill savings of ALEX RL across episodes, benchmarked against ALEX DP. The shaded area represents the variance between runs.

As evident from Figure~\ref{fig: Avg Building Return}, ALEX RL manages to achieve bill savings that slightly exceed those of ALEX DP. It is crucial to note that ALEX DP performs its search for one day ahead, while ALEX RL is not constrained in the duration of its load shifting. These results indicate that, for the CityLearn 2022 community, there is ample opportunity to shift load over several days.

\begin{figure*}[h]
    \centering
\includegraphics[width=0.85\textwidth]{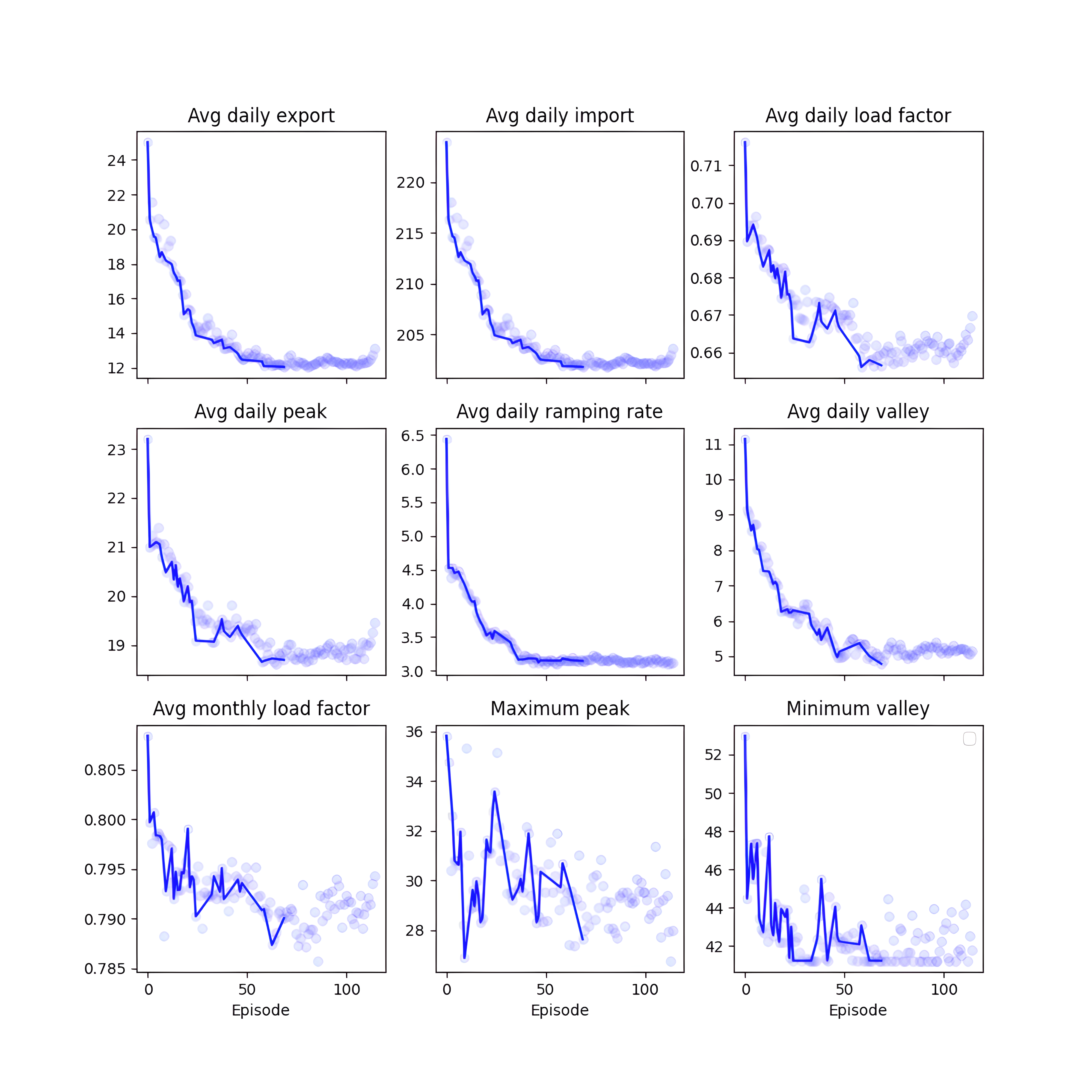}
    \caption{Performance of recorded community-level metrics per episode throughout training. The opaque scattered data points represent singular episode equivalents, while the blue line depicts the metric performance of the most recent highest return achieved.}
    \label{fig: metrics_during_learning}
\end{figure*}

To strengthen the correlation between achieved bill savings and evaluation metrics, Figure~\ref{fig: metrics_during_learning} tracks the performance of the median-performing run of ALEX RL in terms of performance evaluation metrics throughout training. A discernible downward trend is evident for all performance metrics, signifying a clear correlation between selfish bill minimization and the selected set of performance metrics.

Qunatitative analysis, summarized in Table~\ref{tab:Correlations}, consistently supports correlations between performance metrics and participant bill savings. These findings affirm that training DRL agents within ALEX incentivize behavior conducive to the emergent suppression of variability in community net load.
\begin{table*}[h]
    \centering
    \begin{tabular}{ll | c | c | c}
         Metric Correlated to & & Episodic Return & Maximum Return \\
            Average daily import [kWh] & $\overline{E_{d, +}}$  & -0.993 (-0.994) & \textbf{-0.994 (-0.995)} \\
            Average daily export [kWh] & $\overline{E_{d, -}}$ & -0.993 (-0.993) & \textbf{-0.994 (-0.994)} \\
            Average daily peak [kW] & $\overline{P_{d,+}}$ & -0.980 (-0.982) & \textbf{-0.982 (-0.982)} \\
            Average daily valley [kW] & $\overline{P_{d,-}}$ & -0.966 (-0.964)& \textbf{-0.975 (-0.972)} \\
            Minimum peak [kW] & $P_{+}$ & -0.466 (-0.470) & \textbf{-0.478 (-0.480)} \\
            Maximum valley [kW] & $P_{-}$ & -0.734 (-0.736) & \textbf{-0.775 (-0.770)} \\
            Average daily ramping rate [kW] & $\overline{R_{d}}$ & -0.934 (-0.932) & \textbf{-0.952 (-0.955)} \\
            Average daily load factor & $1-L_{d}$ & -0.726 (-0.730)&  \textbf{-0.833 (-0.833)} \\
            Average monthly load factor & $1-L_{m}$ & -0.982 (-0.980) & \textbf{-0.985 (-0.982)} \\
    \end{tabular}
    \caption{Pearson's correlations between the metrics and achieved bill savings; the rightmost column correlates Maximum Return episodes and their respective metric performance, while the middle column correlates episodic return and the respective episodic metric performance; the bracketed number denotes the average correlation over 5 training runs, whereas the non-bracketed number denotes the correlation of the run achieving the median return.}
    \label{tab:Correlations}
\end{table*}

These results strongly suggest that the observed correlations between performance metrics and return are consistent across runs. Furthermore, the observed maximum return correlations are consistently higher than the episodic equivalent. Considering ALEX's nature as a mixed-form stochastic game, this outcome is not necessarily surprising and might result from the convergence path towards a Nash equilibrium. This implies that episodes with higher returns tend to be episodes where the agent policies are closer to a joint best response scenario.

The performance of the median performing set of DRL agents is compared to the proposed benchmarks in Table~\ref{tab:ResultsTable}.
As a result of significantly enhancing the utilization of locally available energy, both the average daily import ($\overline{E_{d, +}}$) and export ($\overline{E_{d, -}}$) decline by 21.9\% and 84.4\%, respectively. Additionally, emergent peak-shaving behavior leads to a lowering of the average daily peak ($\overline{P_{d,+}}$) and valley ($\overline{P_{d,-}}$) by 27.0\% and 71.1\%, respectively, while the maximum peak ($P_{+}$) and minimum valley ($P_{-}$) also shrink by 16.0\% and 27.0\%, respectively. This behavior also results in the smoothing of moment-to-moment community net-load demand, leading to a 26\% decrease in the ramping rate ($\overline{R_{d}}$) and a mitigation of the overall community net-load swing, which reduces the daily load factor ($1-L_{d}$) and monthly load factor ($1-L_{m}$) by 11.0\% and 3.6\%, respectively. In summary, ALEX RL significantly mitigates the effects of community-level variability across all measured metrics.

\begin{table*}[ht]
    \centering
\resizebox{\textwidth}{!}{
    \begin{tabular}{ll|c|c|c|c}
                           Metric & & NoDERMS      & IndividualDERMS  & ALEX DP &ALEX RL\\
         Average daily import [kWh] & $\overline{E_{d, +}}$     & 258.54            & 214.81            &202.68             &\textbf{201.83} \\
         Average daily export [kWh] & $\overline{E_{d, -}}$     & -77.48            & -26.49            &-12.46             &\textbf{-12.04}   \\
         Average daily peak [kW] & $\overline{P_{d,+}}$        & 25.61             & 19.95             &19.44              &\textbf{18.69}   \\
         Average daily valley [kW] & $\overline{P_{d,-}}$       & -16.55            & -6.35             &\textbf{-1.67}     &-4.78    \\
         Maximum peak [kW] & $P_{+}$                            & 49.06             &42.37              &42.37              &\textbf{41.22}   \\
         Minimum valley [kW] & $P_{-}$                         & -37.86            & -36.8             &-29.34             &\textbf{-27.62}   \\
         Average daily ramping rate [kW] & $\overline{R_{d}}$   & 4.28              & 2.87              &\textbf{2.84}      &3.15   \\
         Average daily load factor & $1-L_{d}$                  & 0.73              & 0.65              &\textbf{0.64}      &0.65   \\
         Average monthly load factor & $1-L_{m}$                & 0.82              & 0.8               &\textbf{0.78}      &0.79   \\
    \end{tabular}
    }
    \caption{Summarized metrics for full simulation on CityLearn2022 data set~\cite{CityLearn2022Data} for NoDERMS, IndividualDERMS and ALEX DP and ALEX DRL scenarios. Values for the NoDERMs, IndividualDERMS and ALEX DP are taken out of May et al. \cite{ALEXV2}.
    Best values are typeset in bold.}
    \label{tab:ResultsTable}
\end{table*}

Further comparative analysis demonstrates the cumulative outperformance of the DRL agents against IndividualDERMS and partial outperformance against ALEX DP. Notably, the ramping rate ($\overline{R_{d}}$) emerges as a sub-performant metric for ALEX RL compared to IndividualDERMS and ALEX DP. Additionally, it is noteworthy that the average daily valley metric ($\overline{P_{d,-}}$) for ALEX RL is significantly higher than ALEX DP, which is somewhat unexpected. While IndividualDERMS and ALEX DP search over a perfect forecast, ALEX RL does not have access to future information and must internally perform some degree of participant net load modeling.
As the most short-term volatility-focused metric, the ramping rate ($\overline{R_{d}}$) is also most sensitive to such misadjustments. The relative disparity in average daily valley ($\overline{P_{d,-}}$) between ALEX RL and ALEX DP may result from a strategic tradeoff, where it is economically safer for the DRL agents to err on the side of selling to the grid than buying from it in the face of an imperfect model. Such a scenario could occur when the market receives significantly more bids than asks in terms of quantity, as the remaining residual load will be settled according to a net-billing scenario.

These results further suggest that ALEX RL compensates for its lack of perfect internal modeling by leveraging its capability to load shift over a longer duration than ALEX DP, resulting in a further decrease in the maximum peak ($P_{+}$) and minimum valley ($P_{-}$). Therefore, ALEX RL's relative outperformance in terms of bill savings does not necessarily translate to a strict outperformance of ALEX DP in terms of evaluation metrics. Overall, ALEX RL's performance closely aligns with ALEX DP, indicating similar levels of emergent, community-level coordination of DERs. The collective results compellingly demonstrate emergent, community-level variability reduction facilitated by automated participation via DRL agents within a LEM, effectively closing the identified research gap.

In summary, the findings underscore the effectiveness of leveraging DRL agents in LEMs for load optimization. This emphasizes the potential for mitigating variability and optimizing energy consumption at a community level.
The presented study narrowly focuses on evaluating the performance of a set of DRL agents within ALEX on a specific dataset. The established framework provides a decentralized, data-driven foundation that could potentially scale to larger, more diverse energy communities. However, the investigation of scaling this framework to broader applications may introduce additional challenges and, therefore, falls outside the scope of this work.

\section{Conclusion}\label{sec: Conclusion}
This study explores the automation of participation in economy-driven LEMs through DRL agents.

The rapid proliferation of DERs at the grid edge increases variability and variance in community net load and challenges electricity grid operations. TE-based DR, facilitated by community LEMs, is emerging as as a viable solution by aligning the interests of electricity end-users and grid stakeholders~\cite{Mengelkamp2019Review, CAPPER_LEM_Review, Dudjak_LEM_Review}. At the same time, insights from DR system pilots highlight the importance of automation~\cite{FedEnergyRegulatoryCommission_DRReport, CHEN2017}.
In response to the decentralized and distributed nature of this challenge, model-free control approaches, particularly DRL, have emerged as promising candidates for the automation of participation in LEMs~\cite{Xu_LEM_Article, Zhou_LEM_Article, Zang_LEM_Article, Chen_LEMDRL_Article, Ye_LEM_2, Ye_LEM_1}. While such prior research predominantly focused on socioeconomic metrics and community net load consumption, there remains a gap in demonstrating a clear reduction in variability or variance.

This article addresses the research gap by utilizing a shared experience~\cite{SharedExperienceAC}, recurrent PPO~\cite{SB3} algorithm with several modifications~\cite{Ilyas_LookAtDRL_PG, Kapturowski_R2D2, HiddenStateRecalc} to train a set of DRL agents within the context of ALEX, an economy-driven LEM where participants aim to selfishly minimize bills without information sharing~\cite{ALEXV1}.The trained DRL agents are compared against benchmark approaches, including a building-level DR strategy and a near-optimal dynamic programming-based solution~\cite{ALEXV2}. Performance is evaluated using a set of metrics capturing net load variance across multiple time horizons, encompassing ramping rate, daily and monthly load factor, peak and average daily import and export. The experiments reveal a clear correlation between relative bill reduction and improvements in the investigated metrics. The trained DRL agents demonstrate promising performance, nearing and, in some instances, surpassing the benchmarks set by the near-optimal approach based on dynamic programming, while consistently outperforming the building-level DR strategy.

Future research directions should focus on exploring the transferability, generalization, and scalability of the presented framework. This includes establishing a clearer performance ceiling by designing more sophisticated DRL algorithms explicitly tailored to the mixed-form stochastic game nature of LEMs like ALEX. 
In this contribution relies on RL agents to learn to maximize reward. However, it would be possible to use the Constrained MDP formulation~\cite{altman99} to help ensure constraints are not violated. Other approaches include safety-aware RL \cite{Hans08} and using a controller to intervene (e.g., Learning from Intervention \cite{DBLP:journals/corr/SaundersSSE17}) if the agent selects a dangerous action.
Additionally, investigating the framework's scalability should address challenges such as increased computational demands and coordination across larger agent networks, as well as exploring interactions between multiple ALEX systems under varying grid conditions. Along these lines, ALEX’s capability to integrate additional types of DERs, such as electric vehicles and renewable energy sources like wind, should also be explored, potentially leading to broader environmental benefits such as reduced emissions. The transfer of the resulting technology platform will also need to consider the diverse landscape of existing grid stakeholder regulations, alongside potential social impacts.

Additionally, extending this study's benchmarking effort to diverse LEM designs and different agent heuristics could offer insights into the factors influencing the efficacy of incentivizing desired behaviors within these systems~\cite{Kiedanski_LEM_Lit_Issues, Papadaskalopoulos_LEM_Lit_Issues}.

\appendix
\section{Auxiliary Information regarding Algorithm Setup}\label{Appendix: Learning Curves}
This appendix aims to enhance the reproducibility of the presented results by providing hyperparameters while also detailing the general approach taken in designing the DRL algorithm and testing the modifications.

The algorithm employed in this study is rooted in the publicly accessible Recurrent PPO implementation from Stable Baselines3 (SB3)\cite{SB3}.
The hyperparameter values that deviate from SB3's recurrent PPO default settings are as follows:
\begin{itemize}
    \item The neural network architecture for both critic and actor consisted of 2 LSTM layers with 256 neurons each, followed by a 64-neuron head, along with a shared 64-neuron feature encoder.
    \item The actor's log standard deviation is initialized as -10 instead of the default 0.
    \item An exponentially decaying learning rate schedule is employed, reducing the learning rate by a factor of 0.69 every 1 million steps.
    \item The size of one mini-batch is set to 72, equivalent to one 3-day trajectory, based on SB3's recurrent PPO implementation for sample collection.
    \item The replay buffer stored 3672 transitions, equivalent to 9 days at 24 steps per day for 17 houses.
    \item The burn-in period for a single sample is set at 50\% of the sample's length, or 36 steps.
\end{itemize}

The algorithm adaptations, design, and hyperparameter choices underwent testing across increasingly complex versions of the experiments discussed in the main body of this article until the performance detailed in the discussion section was achieved.
The testing progression began with artificial load profiles, aiming to optimize net billing, then advanced to optimizing net billing on the City Learn dataset for a singular month, then the full year, and finally transitioned to the target application.The advantage of this iterative process lies in the clearly defined optimal returns for the test scenarios.

The observation space is the result of a deliberate addition of features that notably improved agent performance across these tests.

PPO algorithms across implementations vary, as discussed by Huang et al~\cite{Huang_PPOReimplementation}. This can be extended to PPO, as the exact nature of making PPO recurrent is up to interpretation. We refer to Pleines et al.~\cite{RecurrentPPO_ImplementationDetails} for an investigation into the characteristics and sensitivities of recurrent PPO.
Tests commenced with the default SB3 recurrent PPO in a shared experience replay setting~\cite{SharedExperienceAC}, followed by the implementation of R2D2~\cite{Kapturowski_R2D2}, then state recalculation~\cite{HiddenStateRecalc}, and finally incorporating a learning rate schedule~\cite{Ilyas_LookAtDRL_PG}. 
Each implementation underwent testing over a small range of hyperparameters for three runs each to ensure consistency, leading to the crystallization of the hyperparameter set used in this study.

The quality of a run was primarily evaluated based on its achieved return, supplemented by the investigation of various RL agent performance metrics.
These metrics, inspired by those discussed in the SB3 documentation and Huang et al.'s insightful blog~\cite{Huang_PPOReimplementation}, encompassed explained variance, KL-divergence, and entropy loss curves.
Even if an algorithm change did not directly impact the agent's average return, it was considered an improvement if, for example, it led to higher explained variance and thereby a stronger critic.

This iterative practice enabled the authors to initiate algorithm development in smaller, constrained versions of the final application, gradually scaling the difficulty of the experiments as the algorithm matured.
Consequently, the algorithm utilized in this study is relatively basic and does not entail a vast array of modifications, focusing instead on targeted adaptations aimed at enabling the agents to construct a robust temporal state representation.

The training of the DRL agents was conducted on a machine with the following specifications:
\begin{itemize}
    \item CPU: Ryzen 5950X
    \item GPU: Nvidia 3090
    \item RAM: 64 GB
    \item Storage: 2TB NVMe SSD
\end{itemize}
The codebase was optimized for Python 11, with the dataset stored in a SQL database for future scalability. A single run of parallel workers learning for 120 episodes took approximately 7 hours. However, it is important to note that runtimes may vary significantly depending on hardware configurations, software environments, and other factors.

\section*{Acknowledgment}

This research has been supported by the Natural Sciences and Engineering Research Council (NSERC) of Canada grant RGPIN-2017-05866 and by the  NSERC/Alberta Innovates grant ALLRP 561116-20. 
Part of this work has taken place in the Intelligent Robot Learning (IRL) Lab at the University of Alberta, which is supported in part by research grants from the Alberta Machine Intelligence Institute (Amii); a Canada CIFAR AI Chair, Amii; Digital Research Alliance of Canada; Huawei; Mitacs; and NSERC.

\bibliographystyle{elsarticle-num}
\bibliography{paper}

\appendix

\end{document}